\documentclass[twocolumn,floatfix,aps,showpacs]{revtex4}

\usepackage{graphicx}

\begin{document}

\title{Exciton spin relaxation in single semiconductor quantum dots}

\author{E.~Tsitsishvili\cite{ad} and R.~v.~Baltz}
\affiliation{Institut f\"ur Theorie der Kondensierten Materie,
             Universit\"at Karlsruhe, D-76128 Karlsruhe, Germany}
\author{H.~Kalt}
\affiliation{Institut f\"ur Angewandte Physik,
             Universit\"at Karlsruhe,  D-76128 Karlsruhe, Germany}

\date{\today}

\begin{abstract}
We study the relaxation of the exciton spin (longitudinal
relaxation time $T_{1}$) in single asymmetrical quantum dots due
to an interplay of the short--range exchange interaction and
acoustic phonon deformation. The calculated relaxation rates are
found to depend strongly on the dot size, magnetic field and
temperature. For typical quantum dots and temperatures below 100
K, the zero--magnetic field relaxation times are long compared to
the exciton lifetime, yet they are strongly reduced in high
magnetic fields. We discuss explicitly quantum dots based on
(In,Ga)As and (Cd,Zn)Se semiconductor compounds.
\end{abstract}

\pacs{78.67.Hc, 72.25.Rb,  63.20.Ls}

\maketitle

The current interest in the manipulation of spins in
semiconductors is based on the ability to control and maintain
spin coherence over practical length and time scales. Of
particular interest for possible applications in quantum computing
is the storage of spins in zero--dimensional semiconductor
structures. Recent theoretical investigations of the single
carrier--spin relaxation in semiconductor quantum dots (QDs) found
that the electron spin--flip transitions are very slow with
relaxation times of milliseconds and longer\cite{Kha,Woods}. But
since most concepts for quantum information processing involving
QDs are based on optical generation, manipulation and read-out of
spins, one has to investigate the dynamics of the excitonic spin.
In our previous communication\cite{TBK} (motivated by experimental
results for InAs QDs\cite{Pai} at high temperatures $T\gtrsim 40$
K), we studied the exciton--spin relaxation in QDs related to a
second--order process governed by optical phonons. Further
detailed theoretical studies of the exciton--spin relaxation in
QDs are still missing.

Experimental investigations on the exciton--spin dynamics in QDs
refer mostly to the spin--coherence
problem\cite{Gupta,Woggon,Hvam} i.e., they determine the
transverse relaxation time $T_{2}$. Studies of the longitudinal
spin--relaxation time $T_{1}$ are rare since they require strict
resonant excitation conditions and/or high magnetic fields. Recent
resonant--excitation experiments on InAs QDs indicate no
exciton--spin relaxation at low temperatures: the exciton--spin is
totally frozen during the radiative lifetime even for high
magnetic fields up to 8 T\cite{Pai}. Similar results are reported
for CdSe QDs with large lateral dimensions (comparable to twice
the exciton Bohr radius of $\sim$ 5 nm in bulk CdSe
crystals)\cite{Flis,Hundt}. However, longitudinal spin--relaxation
times comparable to the exciton lifetime were deduced from high
magnetic field experiments for much smaller CdSe QDs\cite{Res}.
The observed magnetic field dependence (the relaxation rate
strongly increases with magnetic fields) points on an
acoustic--phonon mediated exciton spin--flip as underlying
mechanism. The same conclusion is drawn from the linear
temperature dependence of the spin--relaxation
rate reported for non--resonant excitation in InGaAs 
disks\cite{Gotoh}.

We propose in this paper an intrinsic mechanism for excitonic
spin--flip transitions at low temperatures resulting from a
deformation--induced exchange interaction. Our model calculations
are able to qualitatively reproduce the above mentioned trends of
the experiments. We study the dependence of the exciton spin
relaxation time $T_{1}$ on temperature, magnetic field, and the 
QD size. Finally, we draw conclusions on the suitability of 
various materials for the storage of optically generated spins.

\begin{figure}[t]
\centerline{\includegraphics[width=6cm]{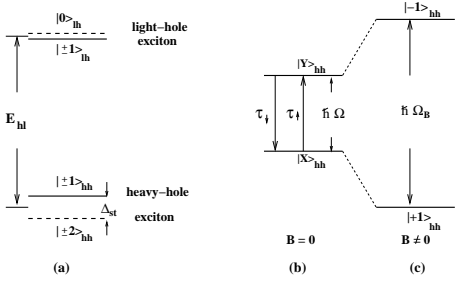}}

\caption{Schematic diagram showing (a) the sublevels of the
heavy--hole--exciton and light--hole--exciton ground states in
symmetrical QDs, (b) the phonon-assisted spin--flip processes
within the radiative doublet of the heavy--hole--exciton in an
asymmetrical QD, and (c) the Zeeman radiative doublet of the
heavy--hole--exciton. Optically inactive states are shown by
dashed lines.\label{f1}}
\end{figure}

Extensive experimental studies have identified the main features
of the exciton fine structure in self--organized QDs by means of
single--dot spectroscopy\cite{Bayer,BK}. Such QDs are usually
strained and have an asymmetrical shape with a height smaller than
the base size. It has been shown that a reduction of the QD
symmetry lifts degeneracies among the exciton states and results,
in particular, in a splitting of the exciton ground state. Thus,
as a consequence of strain and confinement in the growth direction
of a QD, the ground states of the heavy--hole (hh) and light--hole
(lh) excitons are well-separated and the hh--exciton has the
lowest energy. The hh-- and lh--exciton quartets are characterized
by the projections $J_{z} = \pm 1$, $\pm 2$ and $J_{z} = \pm 1$,
$0$ of  the total angular momenta $J = 1, 2$, respectively. The
short-range exchange interaction splits the ground states of both
hh-- and lh--excitons into doublet states (so--called
singlet--triplet splitting), as is shown in  Fig.~\ref{f1}(a). The
lateral anisotropy of a QD leads to a further splitting of the
opticaly allowed doublets $|\pm 1\rangle$ into two levels (labeled
$|X\rangle$ and $|Y\rangle$) with dipole moments along the two
nonequivalent in--plane QD axes. This splitting is due mostly to
the long--range exchange interaction and originates from the
lateral elongation of the QD's\cite{Bayer,Iv1,Ivchenko}.
Continuous wave single--dot spectroscopy experiments have clearly
evidenced the two related, linearly polarized optical transitions
e.g., in GaAs interfacial dots \cite{Gm1} or self-organized $\rm
InGaAs$ QD's\cite{BK}. Measured magnitudes of this splitting reach
some tens or even hundreds of $\mu$eV\cite{Pai,BK,Gm1}. Relaxation
processes between the radiative states changes the occupation of
the exciton levels and are manifested in a change of the linear
degree of polarization of the luminescence\cite{Pai,TBK}.

In what follows we consider (longitudinal) spin--relaxation processes between 
the allowed $|X\rangle$-- and $|Y\rangle$-- states of the hh--exciton.
In such processes, the hh--exciton stays in the same 
spatial state and just flips its spin (i.e., the electron and the hole spin 
flip simultaneously), by emitting or absorbing an acoustic phonon.
Since  the bare electron/hole--phonon interaction does not contain spin 
operators, it cannot directly couple the involved exciton states,
however, this can occur via the hh and lh exciton mixing due to 
the interplay of the the short--range exchange interaction and the 
lattice deformations.

The short--range (isotropic) exchange interaction is given by the Hamiltonian
\cite{BP}
\begin{eqnarray}
{\mathcal{H}}_{ex} = - \frac{2}{3}\;\Delta_{st} \;\vec{\sigma}\;\vec{J}, 
\label{sH}
\end{eqnarray}
where $\vec{\sigma}$ and $\vec{J}$ are the electron spin and the hole
total angular momentum 
operator, respectively, and $\Delta_{st}$ is the (singlet--triplet) 
exchange energy. Phonon--induced deformations come into play
via the off--diagonal terms in the Bir--Pikus Hamiltonian\cite{BP}
\begin{eqnarray}
{\mathcal{H}}(\varepsilon) = b\;\sum_{i}J_{i}^2\;\Bigl(\varepsilon_{ii}
- \frac{1}{3}\;\varepsilon\Bigr) +
\frac{2}{\sqrt{3}}\;d \sum_{i>j}\Bigl[J_{i} J_{j}\Bigr]\;\varepsilon_{ij},
\label{hj}
\end{eqnarray}
where b and d are the exciton deformation potentials,
$\varepsilon_{ij}$ is the
deformation tensor, and  $[J_{i} J_{j}] = J_{i} J_{j}+J_{j} J_{i}$.
(An overall shift of $-a\varepsilon$ of the lh and hh exciton levels is
omitted).

In the basis of the hh-- and lh--exciton states
($|-1\rangle_{hh}$, $|+1\rangle_{hh}$ and $|-1\rangle_{lh}$,
$|+1\rangle_{lh}$, respectively), the total Hamiltonian
${\mathcal{H}} = {\mathcal{H}}(\epsilon) + {\mathcal{H}}_{ex}$ is
represented by the following matrix
\begin{eqnarray}
\mathcal{H} & = &  \left|
\begin{array}{cccc}
0 & 0 & \frac{\Delta_{st}}{\sqrt{3}} & j^{*}\\
0 & 0  & j & \frac{\Delta_{st}}{\sqrt{3}}\\
\frac{\Delta_{st}}{\sqrt{3}} & j^{*} & E_{lh} & 0\\
j & \frac{\Delta_{st}}{\sqrt{3}} & 0 &  E_{lh}\\
\end{array}
 \right|.
\label{HELE}
\end{eqnarray}
The origin of energy axis is fixed at the hh--exciton states which
are separated from the lh-- exciton by $E_{lh}$, see
Fig.~\ref{f1}(a). The deformation--dependent off--diagonal term is
$j = \sqrt{3} b (\varepsilon_{xx} - \varepsilon_{yy})/2 - id
\varepsilon_{xy}$. Here, we neglect the anisotropic part of the
splitting of the allowed doublets, since the energy difference
$E_{lh}$ is, typically, much larger (of the order of several tens
of a meV) than the fine structure energies\cite{Bayer}. For the
same reason, the hh--lh--exciton mixing  can safely be regarded as
a perturbation.  It follows from Eq. (\ref{HELE}) that
perturbations which mix hh-- and lh--exciton states result in a
coupling of the $|-1\rangle_{hh}$-- and $|+1\rangle_{hh}$--states
and, consequently, lead to the mixing of the $|X\rangle$-- and
$|Y\rangle$-- states of the hh--exciton. For QDs which are
elongated in the [110] direction, this coupling is due to
vibrations causing $(\varepsilon_{xx} - \varepsilon_{yy})$
deformations
\begin{eqnarray}
\langle X|{\mathcal{H}}|Y \rangle = - \imath \; \frac{(j + j^{*})\;\Delta_{st}}
{\sqrt{3} E_{lh}} \equiv - \imath \; \frac{\Delta_{st}}{E_{lh}}\;b\;
(\varepsilon_{xx} - \varepsilon_{yy}).
\label{hh}
\end{eqnarray}
Thus, a coupling of the $|X\rangle$-- and $|Y\rangle$--states is a
result of an interplay between the short-range exchange
interaction and deformations: neither the exchange interaction nor
the deformation perturbation $j$ alone can couple these states.
According to Eq.~(\ref{hh}), the matrix element for the relaxation
processes among the $|X\rangle$-- and $|Y\rangle$-- states of the
hh--exciton is
\begin{eqnarray}
M_{XY} =  - \imath \; \frac{\Delta_{st}}{E_{lh}}\;b\;
\langle (\varepsilon_{xx} - \varepsilon_{yy}) \rangle,
\label{ME}
\end{eqnarray}
where the expectation value is taken with respect to the spatial part of the
exciton wave function.
The elastic strain is given by
\begin{eqnarray}
\varepsilon_{ij} =
\frac{1}{2}\;\Bigl(\;\frac{\partial u_{i}}{\partial r_{j}} +
\frac{\partial u_{j}}{\partial r_{i}}\;\Bigr),
\label{df}
\end{eqnarray}
where the phonon displacement field is
\begin{eqnarray}
\vec{u} = \frac{1}{\sqrt{V}}\;\vec{e}(\vec{q})\;\sqrt{\frac{\hbar}{2 \rho
\omega_{q}}}\; \Bigl(e^{i \;
\vec{q}\;\vec{r}} \; a_{\vec{q}} + e^{-i \; \vec{q}\;\vec{r}} \;
a_{\vec{q}}^{+}\Bigr).
\label{ds}
\end{eqnarray}
Here $\vec{q}\;$,   $\vec{e}$ and $\omega_{q} = s q$ are
the phonon momentum ($q = |\vec{q}|$),
polarization vector and frequency, respectively,
V is the normalization volume,
$\rho$ is the mass density, and $s$ is the sound velocity.
Upon substitutions given by Eqs.\ (\ref{df}) and (\ref{ds}) we obtain
for the matrix element in Eq.~(\ref{ME})
\begin{eqnarray}
|M_{XY}|^2&=&\frac{1}{V} \Bigl(\frac{\Delta_{st}}{E_{lh}}\Bigr)^2
\frac{\hbar b^2}{2 \rho \omega_{q}}
\;(q_{x} e_{x} - q_{y} e_{y})^2 \; |M_{or}|^2,\label{OME}\\
M_{or}(\vec{q}) &=& \langle \Phi_{0}|e^{i\; \vec{q}\;\vec{r_{e}}} +
e^{i\;\vec{q}\;\vec{r_{h}}}| \Phi_{0}\rangle ,
\label{orb}
\end{eqnarray}
where indices $e$ and $h$ stand for electrons and holes,
$M_{or}$ is the orbital part of the matrix element, and
$ \Phi_{0}(\vec{r_{e}},\vec{r_{h}})$ is the exciton
ground state envelope wave function.

The spin--relaxation rates accompanied by phonon emission and absorption
are given by Fermi's golden rule
\begin{eqnarray}
\frac{1}{\tau_{\downarrow}} &=& \frac{2 \pi}{\hbar} \:\sum_{\vec{q}}\:
\sum_{\vec{e}}\;
 |M_{XY}|^2\;\delta(\hbar \Omega - \hbar \omega_{q})\;
(N_{\omega_{q}} + 1), \nonumber\\
\frac{1}{\tau_{\uparrow}} &=&
\frac{2 \pi}{\hbar} \:\sum_{\vec{q}}\:\sum_{\vec{e}}\;
 |M_{XY}|^2\;\delta(\hbar \Omega - \hbar \omega_{q})\; N_{\omega_{q}},
\label{dr}
\end{eqnarray}
where $\hbar \Omega$ is the energy splitting between  the
$|X\rangle$-- and $|Y\rangle$--states (see Fig.~\ref{f1}(b)), and
$ N_{\omega_{q}} = 1/(e^{\hbar \omega_{q}/k_{B}T} -1)$ is the
thermal phonon distribution function.

We discuss now what determines the value of the orbital part of
the matrix element Eq.~(\ref{orb}). Because of energy
conservation,  $|M_{or}| \equiv |M_{or}^{e} + M_{or}^{h}|$ has
to be calculated for the phonon momentum $q = \omega_{q}/s \equiv
\Omega/s$, while the characteristic extent of the orbital wave
function $\Phi_{0}$ is, roughly speaking, the lateral size $L$ of
the QD. (For flat QDs, the QD hight $L_{z} \ll L$.) Due to the
oscillatory behavior of the exponential functions in $M_{or}$, we
have $|M_{or}^{e,h}| \approx 1$, if $ \Omega L/s \lesssim 1$. To
illustrate this approximation, we consider the model of a strongly
confined QD with the shape of an parallelepiped and infinite
barrier. Then the exciton states are reasonably well approximated
by noninteracting electron--hole pair states. In addition, the
single carrier--phonon matrix element
$|M_{or}^{e}(\vec{q})|=|M_{or}^{h}(\vec{q})|\equiv|M_{or}(\vec{q})|$
separates in the $x, y, z$ coordinates. For a confined direction
(say, the $x$--direction), the following expression holds
\begin{eqnarray}
|M_{or}(q_{x})| = \left| \frac{\sin Q_{x}}{Q_{x}} + \frac{Q_{x} \sin Q_{x}}
{Q_{x}^2 - \pi^2}\right|,
\label{xd}
\end{eqnarray}
where $Q_{x}=q_{x} L_{x}/2$ and $L_{x}$ is the QD size in the
$x$--direction. As seen from Eq.~(\ref{xd}), the matrix element
$|M_{or}(q_{x})|$ is smaller than unity for any $q_{x} \neq 0$,
but it approaches unity for small $q_{x} L_{x} \ll 1$ (
$|M_{or}(q_{x})| \sim |1 - Q_{x}^2/\pi^2|$). With increasing
$q_{x} L_{x} \gg 1$, $|M_{or}(q_{x})|$ decreases rapidly (
$|M_{or}(q_{x})| \sim |\sin Q_{x}/Q_{x}|$). Thus, if we replace
the orbital matrix element in Eq.~(\ref{OME}) by unity,
$|M_{or}(\vec{q})|^2 \equiv 1$, the exciton--spin relaxation rates
Eq.~(\ref{dr}) will be overestimated. Note that for typical values
of $\Omega $ and $s$ we have $(s/\Omega) \sim 40$ nm, $L$ is
usually about $10$--$20$ nm\cite{tv}, so that the above mentioned
assumption $ \Omega L/s \lesssim 1$ is reasonable.
As a result, a lower limit for the relaxation times is given by
\begin{eqnarray}
\tau_{\downarrow} &=& \frac{9}{4}\; \frac{\hbar \rho s^2}{b^2}\;
\Bigl(\frac{E_{lh}}{\Delta_{st}}\Bigr)^2 \;
\Bigl(\frac{s}{\Omega}\Bigr)^{3} \; \Bigl(1 - e^{- \hbar
\Omega/k_{B}T}\Bigr),
\label{sdt}\\
\tau_{\uparrow} &=&
\tau_{\downarrow} \;  e^{\hbar \Omega/k_{B}T}. \label{sut}
\end{eqnarray}
In order to obtain a numerical estimate, we use typical
parameters\cite{tv}, together with an estimated hh--lh--exciton
splitting of $E_{lh} = 10$ meV, i.e., $E_{lh}/\Delta_{st} = 50$.
As a result, we get for the spin relaxation time scale $T_{1}
\equiv (\tau_{\downarrow}^{-1}+\tau_{\uparrow}^{-1})^{-1} \sim
1500$ ns (at $T = 10$ K)  and  $T_{1} \sim 150$ ns  (at $T = 100$
K). Thus, {\it for typical quantum dots and low temperatures, the
zero--magnetic field relaxation times are very long compared to
the exciton lifetimes of  $\sim 1$ ns}. This is in qualitative
agreement with resonant--excitation experiments at low
temperatures\cite{Pai,Flis}. Please note, that because of the
small interlevel splitting $\hbar \Omega \sim 0.1$ meV, the
relaxation times $\tau_{\downarrow}$ and $\tau_{\uparrow}$ are
already comparable at $T \gtrsim 2$ K. For higher temperatures
$T_{1}$ is inversely proportional to the temperature. But for $T
\gtrsim 100$ K, the interaction with LO phonons via a
second--order process will be the dominant relaxation
mechanism\cite{Pai,TBK}.

The calculated spin relaxation times according to
Eqs.~(\ref{sdt},\ref{sut}) are large partly because of the small
phonon density of the states (which is $\sim (\Omega/s)^2$) at the
scale of the interlevel splitting $\hbar \Omega$. But,
considerably large interlevel splittings of $\hbar \Omega_{B} = g
\mu_{B} B \sim 1.5$ meV (where $g$ is the exciton effective
$g$--factor) arise in high--field magneto--optical experiments.
Here an enhancement of spin--relaxation rate is expected. The spin
flip in Faraday geometry ($\vec{B}\parallel z$) occurs between the
$|-1\rangle_{hh}$-- and $|+1\rangle_{hh}$-- states of the
hh--exciton (see Fig.~\ref{f1}(c))\cite{Bayer}. Consequently,
exciton--spin relaxation processes between the Zeeman sublevels
are reflected in a change of the circular degree of polarization
of the luminescence.

For magnetic fields and zero temperature, a lower limit of the spin 
relaxation time is given by
\begin{eqnarray}
T_{1}(B) & \simeq & \tau_{\downarrow}(B) =
\frac{9}{4}\; \frac{\hbar \rho s^2}{\tilde b^2}\;
\Bigl(\frac{E_{lh}}{\Delta_{st}}\Bigr)^2\;
\Bigl(\frac{s}{\tilde \Omega_{B}}\Bigr)^{3},
\label{sB}\\
\tilde b^2 &=& b^2 + \frac{\Omega_{B}}{3 \sqrt{\Omega_{B}^2 + 
4 \Omega^2}} d^2, 
\label{sB1}\\
\tilde \Omega_{B} &=& \Omega + \Omega_{B}.
\label{sdtB}
\end{eqnarray}
The relaxation time in Eq.~(\ref{sdtB}) varies as the square of 
the ratio
$E_{lh}/\Delta_{st}$ and depends strongly on the QD size because
both the splitting $E_{lh}$ and the exchange energy $\Delta_{st}$
are size--dependent\cite{Tag,rem}. If the splitting $E_{lh}$ is
due mostly to the strain (large lattice mismatch), $T_{1}$ changes
with the lateral  and vertical size as $L^{4}$ and  $L_{z}^{2}$,
respectively. For small lattice mismatch, $E_{lh}$ is due mostly
to the vertical confinement and, therefore, $T_{1} \sim L^{4}
L_{z}^{-2}$. In any case, {\it the relaxation time $T_{1}$
strongly decreases when decreasing the lateral size $L$ of a QD}.
This result can explain qualitatively the observations for the
CdSe QDs of different lateral sizes in high magnetic fields. As
noted above, no spin relaxation was found for {\it large} CdSe QDs
(with $L$ about 10 nm)\cite{Hundt}, while for {\it small} CdSe QDs
(with $L$ about 3 nm - 5 nm) an efficient exciton--spin relaxation
was observed \cite{Res}.

To obtain numerical estimates for the spin relaxation in magnetic
fields, we use the experimental values for $E_{lh}, \Delta_{st}$
and $g$, and we take the typical values\cite{tv} for other factors
in Eq.~(\ref{sB}). For InAs/GaAs QDs, we use $\Delta_{st} =
0.2$ meV,  $g = 3$\cite{BK} and $E_{lh} = 30$ meV ($E_{lh}$ of
several tens of a meV is reported in Ref.\cite{Bayer}). As a
result we get a lower limit for the relaxation time $T_{1} \simeq
13/B^{3}$ ($\mu$s T$^{3}$) which is much larger than the exciton
lifetime scale $\sim 1$ ns even at high $B$ (e.g., $T_{1} \simeq
25$ ns at $B = 8$ T). Indeed, no spin relaxation is found for low
temperature experiments in high magnetic fields \cite{Pai}.

This case is different for small CdSe/ZnSe QDs investigated in
Ref.\cite{Res,Verbin}. Here, values of $\Delta_{st} \sim 1$ meV,
$g \sim 2$ and $E_{lh} \sim 40$ meV are reported. As a result we
calculate $T_{1} \sim 5$ ns at $B = 6$ T
which is close to values of 2.5 ns - 3.4 ns deduced from the
experiments\cite{Res}. For large CdSe/ZnSe QDs investigated in
Ref.\cite{Hundt} no data on the exchange energy $\Delta_{st}$ are
reported but we still can estimate the relaxation rate. The
lateral size of the QDs in this case is about a factor of 2.5
larger than in \cite{Res}. The consequence is a much smaller
exchange splitting $\Delta_{st} \sim 0.2$ meV \cite{rem} which
results in much longer relaxation times of $\sim 100$ ns (at $B =
6$ T). As already mentioned, no spin relaxation is observed in the
experiments of Ref.\cite{Hundt}.

Let us now consider the magnetic field dependence of the
spin--relaxation time which was measured in small CdSe/ZnSe QDs
\cite{Res}. The experimental results indicate a roughly
anti--proportional decrease $T_1\propto B^{-1}$ up to 8 T.
As seen from Eqs.~(\ref{sB},\ref{sdtB}), $T_{1}$ scales 
as $B^{-1}$ for small $B$ ($\Omega_{B} < \Omega$), in agreement 
with the experiment. But it would depend much stronger on the 
magnetic field (as $B^{-3}$) for large $B$. This difference to 
the experiment can originate from the treatment of the
orbital matrix element, which up to now has been fixed as 1. But a
factor of $1/|M_{or}|^2$ has to be accounted for explicitly in
Eq.~(\ref{sB}) for high magnetic fields. The reason is, that the
relevant phonon momentum $q(B) \simeq \Omega_{B}/s$ is linearly
proportional to $B$. Thus, $1/|M_{or}|^2$ increases with magnetic
field and becomes important at $q(B) L \geq 1$. This is the case
e.g., for $B > 5$ T for CdSe QDs with a lateral size of $L \sim 4$
nm and is the regime of the experiments in \cite{Res}. 
So we can confirm the general trend of the experiment\cite{Res},
namely that $T_{1}$ decreases with magnetic field. An analytical
dependence of $T_{1}$ on $B$ (at high $B$) would require an 
explicite calculation of $|M_{or}|^2$ with the QD exact parameters, 
which for the QDs used in Ref.\cite{Res}, unfortunately, are not 
known.

There are many experiments, which determine the polarization
degree of the luminescence from QDs under nonresonant excitation
conditions. We just want to discuss the results of
Ref.\cite{Gotoh} as a typical representative. Rather short
polarization decay times on the order of 1 ns have been found
here. But definite conclusions on the underlying spin--relaxation
mechanisms are not possible in a scenario where the electron--hole
pairs relax through a multitude of barrier states and/or excited
quantum--dot states. To identify spin--relaxation processes, the
splitting of the exciton ground state has to be large (as for large
magnetic fields) or strictly resonant excitation has to be used
(see \cite{Pai}).

The above discussions show that a direct quantitative comparison
of our proposed relaxation model to experimental data is at the
moment difficult at best. One obvious reason is that the
relaxation times, once they are much larger than the exciton
lifetime, cannot be quantified experimentally. The second reason
is, that only very few experiments under strict resonant
excitation and/or in high magnetic fields have been performed. But
still, there are a couple of experimental trends which are
explained consistently by our model. The value and magnetic--field
dependence of $T_1$ in small CdSe QDs can be reproduced. We
correctly predict the experimental findings that long relaxation
times are expected for InAs QDs as well as large CdSe QDs at low
temperatures even in strong magnetic fields.

The experimental differences found for InAs vs CdSe QDs and for
large vs small CdSe QDs strongly support our proposed model. Let's
consider alternative spin--relaxation mechanisms within the
radiative doublet. Processes relying on the exciton motion like
for quantum-well excitons \cite{Maialle} are suppressed in QDs.
Assume, the electron and hole would individually flip their spins.
In contrast to our proposed mechanism, this is a higher--order
process involving intermediate dark exciton states (the excitons
finally have to return to radiative states to contribute to a
depolarized luminescence). The relaxation time here is determined
by the longest spin--flip time for a single carrier (electron or
hole). The spin flip of the electron has to rely on mechanisms
based on the spin--orbit coupling. These relativistic effects have
been shown by Khaetskii and Nazarov \cite{Kha}
to be inefficient in QDs. 
Finally, the spin--orbit interaction and carrier--phonon coupling 
are not different enough in InAs and CdSe or for large and small 
QDs to explain that the spin relaxation times change by orders of
magnitude. This dramatic change found in the experiments is,
however, readily explained by the strongly differing exchange
splitting in InAs and CdSe and its significant enhancement in
small dots. Since the exchange splitting enters quadratically in
our proposed relaxation scheme, we are quite convinced to have
identified the most prominent spin--relaxation process for QDs.

\begin{table*}
\caption{Semiconductor compounds forming QDs (first column) and
matrices (sixth column), $\Delta_{st}^{b}$ and $a_{ex}$ are the
exchange splitting and the exciton Bohr radius in bulk compounds forming QDs,
$\Delta_{st}$ gives typical values of the  exchange splitting in QDs with
sizes of $L_{z} = 3$ nm and $L = 15$ nm.
$m_{0}/m = m_{0} (1/m_{lh}-1/m_{hh})$, where $m_{lh}$ and $m_{hh}$ are the lh--
and hh--masses,  respectively.
$|\Delta a|/a = (a_{d}-a_{m})/0.5(a_{d}+a_{m})$ and
$\Delta E_{g}=E_{gm} - E_{gd}$ with lattice constants $a_{d}$ and $a_{m}$
and energy gaps $E_{gd}$ and  $E_{gm}$ in
the QD and matrix materials, respectively.
Material constants are taken from Ref.~\onlinecite{Landolt}, except
$\Delta_{st}^{b}$ and
$a_{ex}$ for InAs crystals which are taken from Ref.~\onlinecite{Fu}.
For the case of CdSe crystals,  $\Delta_{st}^{b}$,
$a_{ex}$,  $\Delta E_{g}$, and $m_{0}/m$ are given for hexagonal CdSe
crystals.\label{tab1}}
\begin{ruledtabular}
\begin{tabular}{lcccclcc}
 QD  & $a_{ex}$ & $\Delta_{st}^{b}$ & $\Delta_{st}$
& $\displaystyle\frac{m_{0}}{m}$
& matrix  & $\displaystyle\frac{|\Delta a|}{a}$  & $\Delta E_{g}$ \\
& (nm) & ($\mu$eV) & $\;(\mu$eV) &   &  &  &  (eV)  \\ \hline
InAs & 38 & 0.3 & 200 & 37.5 & GaAs &  0.069 & 1.1  \\
InAs & 38 & 0.3 & 200 & 37.5 & Al$_{0.3}$Ga$_{0.7}$As &  0.068 & 1.8  \\
GaAs &11 & 20  & 250 & 9.3 &  Al$_{0.3}$Ga$_{0.7}$As &  0.0005 & 0.48  \\
In$_{0.6}$Ga$_{0.4}$As & 24 & 8.2  & 220 & 17.7 & GaAs & 0.042 & 0.62 \\
In$_{0.4}$Ga$_{0.6}$As & 22 & 12.1  & 230 &  13.8 & Al$_{0.5}$Ga$_{0.5}$As
& 0.027 & 1.22 \\
In$_{0.7}$Ga$_{0.3}$As & 31 & 6.2  & 210 & 20.6 & GaAs & 0.049 & 0.73 \\
CdSe & 5.4 & 130 & 450 & 4.7 & ZnSe & 0.066 & 0.9 \\
\end{tabular}
\end{ruledtabular}
\end{table*}

Finally, we want to discuss different material systems with
respect to their suitability for spin storage. The relevant
parameters for several semiconductor compounds forming QD's and
their matrices are presented in Table~\ref{tab1}. Among the
different material parameters in Eq.~(\ref{sdtB}), the
characteristic energies $E_{lh}$ and $\Delta_{st}$ are strongly
affected by the structural composition. Thus we can only give
general trends. We will consider QDs with typical sizes of about
$L_{z} \sim 2$--$5$ nm and $L \sim 10$--$20$ nm which are in the
strong vertical confinement regime. Moreover,  $\rm
In_{x}Ga_{1-x}$As QDs are also strongly confined in the lateral
plane, and the exchange splitting $\Delta_{st}$ in these QDs
significantly exceeds its bulk value of
$\Delta_{st}^{b}$\cite{rem}. In CdSe QDs the bulk value of the
exchange splitting is already quite large. In CdSe QDs with
lateral sizes larger than the exciton radius of $a_{ex}\sim5$ nm
the exchange splitting is additionally affected by the vertical
confinement. Since the exchange interaction in (In,Ga)As crystals
is much weaker than in CdSe crystals, $\Delta_{st}$ is of the same
order of magnitude\cite{rem1} in strongly confined $\rm
In_{x}Ga_{1-x}$As QDs and large, weakly confined CdSe QDs.
Considering additionally the strain--induced lh--hh splitting
$E_{lh}$, which has to be large to suppress spin relaxation, InAs
QDs have the longest relaxation time $T_1$.

Less suitable for spin storage is the GaAs/Al$_{0.3}$Ga$_{0.7}$As
system since its lattice mismatch is very small. Thus, among the
QD structures listed in Table I, the shortest $T_{1}$ relaxation
times can be expected for strongly confined
GaAs/Al$_{0.3}$Ga$_{0.7}$As QDs. Likewise, small $T_1$ times are
predicted for small CdSe QDs since in the strong lateral
confinement regime, $\Delta_{st}$ is considerably enhanced.

In conclusion, the exciton--spin relaxation within the radiative
doublet of the exciton ground state in single asymmetrical QDs is
studied. As a possible intrinsic mechanism for such a process, the
exciton spin--acoustic phonon coupling via the strain--dependent
short range exchange interaction is proposed. For zero-magnetic
fields and low temperatures, the calculated $T_{1}$--relaxation
times for typical QDs are long compared to the exciton lifetime.
A strong reduction of the relaxation times occurs in QDs in high
magnetic fields. Nevertheless, numerical estimates for InAs QD's
and large CdSe QDs give large values for the $T_{1}$--relaxation
times even for high $B \sim$ 8 T (up to a few of tens of
nanoseconds). In addition, the relaxation time $T_{1}$ strongly
decreases in strongly confined (in the lateral plane) QDs. The
$T_{1}$--relaxation times estimated for small CdSe QDs in high
magnetic fields reduce to
the nanosecond scales. These predictions of our model are in
qualitative agreement with experimental findings. Within the
considered mechanism, we conclude that InAs QDs and CdSe QDs with
typical lateral sizes of $L \sim 10$ -- $20$ nm display a weak
exciton spin relaxation, whereas for typical
GaAs/Al$_{0.3}$Ga$_{0.7}$As QDs and small CdSe QDs the $T_{1}$
times are expected to be rather short.

This work was supported by the Center for Functional
Nanostructures (CFN) of the Deutsche Forschungsgemeinschaft (DFG)
within project A2.



\begin{thebibliography}{99}
\bibitem[*]{ad}
     Permanent address: Institute for Cybernetics, Academy of Science,
     S.~Euli 5, 380086, Georgian Republic.
\bibitem{Kha}
     Alexander V.~Khaetskii and Yuli V.~Nazarov,
     Phys. Rev. B \textbf{64}, 125316 (2001).
\bibitem{Woods}
     L.M.~Woods, T.L.~Reinecke, and Y.~Lyanda--Geller,
     Phys. Rev. B \textbf{66}, 161318(R) (2002).
\bibitem{TBK}
     E.~Tsitsishvili, R.~v.~Baltz, and H.~Kalt,
     Phys. Rev. B \textbf{66}, 161405(R) (2002).
\bibitem{Pai}
     M.~Paillard, X.~Marie, P.~Renucci, T.~Amand, A.~Jbeli,
     and J.M.~Gerard, Phys. Rev. Lett. \textbf{86}, 1634 (2001)
     and references cited therein.
\bibitem{Gupta}
     J.A.~Gupta, D.D.~Awschalom, X.~Peng, and A.P.~Alivisatos,
     Phys. Rev. B \textbf{59}, R10421 (1999).
\bibitem{Woggon}
     P.~Borri, W.~Langbein, S.~Schneider, U.~Woggon,
     R.L.~Sellin, D.~Ouyang, and D.~Bimberg,
     Phys. Rev. Lett. \textbf{87}, 157401 (2001).
\bibitem{Hvam}
     D.~Birkedal, K.~Leosson, and J.M.~Hvam,
     Phys. Rev. Lett. \textbf{87}, 227401 (2001).
\bibitem{Flis}
     T.~Flissikowski, A.~Hundt, M.~Lowisch, M.~Rabe, and F.~Henneberger,
     Phys. Rev. Lett. \textbf{86}, 3172 (2001).
\bibitem{Hundt}
     A.~Hundt, T.~Flissikowski, M.~Lowisch, M.~Rabe, and
     F.~Henneberger, Phys. Status Solidi B \textbf{224}, 159 (2001).
\bibitem{Res}
     Y.~Oka, S.~Permogorov, R.~Pittini, J.X.~Shen, K.~Kayanuma,
     A.~Reznitsky, L.~Tenishev, and S.~Verbin, Physica E \textbf{10},
     315 (2001).
\bibitem{Gotoh}
     H.~Gotoh, H.~Ando, H.~Jamada, A.~Chavez-Pirson, and J.~Temmyo,
     Appl.Phys.Lett. \textbf{72}, 1341 (1998).
\bibitem{Bayer}
     For recent review see
     M.~Bayer, G.~Ortner, O.~Stern, A.~Kuther, A.A.~Gorbunov, A.~Forchel,
     P.~Hawrylak, S.~Fafard, K.~Hinzer, T.L.~Reinecke, S.N.~Walck,
     J.P.~Reithmaier, F.~Klopf, and  F.~Sch\"afer,
      Phys. Rev. B \textbf{65}, 195315 (2002).
\bibitem{BK}
     M.~Bayer, A.~Kuther, A.~Forchel, A.~Gorbunov, V.B.~Timofeev,
     F.~Sch\"afer, J.P.~Reithmaier, T.L.~Reinecke, and S.N.~Walck,
     Phys. Rev. Lett. \textbf{82}, 1748 (1999).
\bibitem{Iv1}
     E.L.~Ivchenko, Phys. Status Solidi A
     \textbf{164}, 487 (1997) and references cited therein.
\bibitem{Ivchenko}
     E.L.~Ivchenko and G.E.~Pikus,
     \textit{Superlattices and Other Heterostructures. Symmetry
     and Optical Phenomena} (Springer-Verlag, Berlin, 1995).
\bibitem{Gm1}
     D.~Gammon,  E.~S.~Snow, B.~V.~Shanabrook, D.~S.~Katzer, and D.~Park,
     Phys. Rev. Lett. \textbf{76}, 3005 (1996).
\bibitem{BP}
     G.L.~Bir and G.E~Pikus, \textit{Symmetry and Strain--Induced Effects
     in Semiconductors} (Wiley, New York, 1975).
\bibitem{tv}
     Typical fine structure energies are:
     $\hbar \Omega \simeq 0.1$meV, $\Delta_{st} = 0.2$ meV.\cite{Bayer}
     Typical material parameters are:
     $s =  5\times 10^{5}$ cm/sec,
     $\rho = 5.3\rm g/cm^{3}$, $b = 2$ eV,  $d = 5$ eV.\cite{Landolt}
\bibitem{Landolt}
     Landolt--B\"ornstein, in \textit{New Series},
     edited by O.~Madelung, M.~Schultz, and H.~Weiss,
     (Springer-Verlag, Berlin, 1982),  Vol. 17b, Group III.
\bibitem{Tag}
     T.~Takagahara, Phys. Rev. B \textbf{47}, 4569 (1993).
\bibitem{rem}
     The exchange splitting in strongly confined QD's can be estimated as
     $\Delta_{st} \sim \Delta_{st}^{b} a_{ex}^{3}/R_{z}R^2$, where
     $R_{z} = L_{z}/2$ and $R = L/2$.
\bibitem{Verbin}
     S.~Verbin, O.Z.~Karimov, A.~Reznitsky, A.A.~Klochikhin, T.~Ruf,
     L.~Tenishev, S.~Permogorov, S.V.~Ivanov, D.~Wolverson, and
     J.J.~Davies, Phys. Status Solidi B
     \textbf{224}, 545 (2001).
\bibitem{Maialle}
     M.Z.~Maialle, E.A.de ~Andrada e ~Silva, and L.J.~Sham,
     Phys. Rev. B \textbf{47}, 15776 (1993).
\bibitem{rem1}
     The exchange interaction for excitons and atoms differs approximately
     by a factor of $(a^{3}/a_{ex}^{3})$\cite{BP} (where $a$ is the lattice
     constant), which results in the similar values of a quantity
     $\Delta_{st}^{b} a_{ex}^{3} \sim 2 \times 10^{4} \mu$eV nm$^{3}$
     for different materials. Consequently,
     $\Delta_{st}$  are close in QD's of different composition but equal
     sizes. For the case of $R > a_{ex}$,
     $\Delta_{st} \sim \Delta_{st}^{b} a_{ex}/R_{z}$.
\bibitem{Fu}
     Huaxiang~Fu, Lin-Wang~Wang, and Alex~Zunger,
     Phys. Rev. B \textbf{59}, 5568 (1999).
\end{thebibliography}
\end{document}